\documentclass[letter,twocolumn]{jpsj3}

\title{Spectral Properties of Interacting One-Dimensional Spinless Fermions}

\author{Masanori Kohno, Mitsuhiro Arikawa$^{1}$, Jun Sato$^{2}$, and Kazumitsu Sakai$^{3}$}
\inst{International Center for Materials Nanoarchitectonics, National Institute for Materials Science, Tsukuba 305-0044, Japan\\
$^{1}$Institute of Physics, University of Tsukuba, Tsukuba 305-8571, Japan\\
$^{2}$Nanotechnology Research Institute, AIST, Tsukuba 305-8568, Japan\\
$^{3}$Institute of Physics, University of Tokyo, Komaba, Tokyo 153-8902, Japan}
\abst{The spectral properties of the spinless fermion model with nearest-neighbor repulsive interactions 
on a one-dimensional lattice 
are investigated using the Bethe ansatz. 
Although its bulk quantities are exactly the same as those of the spin-1/2 XXZ chain, 
the difference in the statistics of particles causes substantial effects on spectral features, such as gapless points of dispersion relations and line shapes of spectral functions. 
In this Letter, we clarify the origin of the differences in spectral features between fermionic and bosonic systems in terms of Bethe ansatz solutions. 
We also confirm that the 2-string solutions have considerable spectral weights in the high-energy regime.}

\kword{Bethe ansatz, spectral functions, fermionic statistics, excitations, one dimension}
\begin{document}
\maketitle

{\it Introduction }- 
In quantum many-body systems, statistics of particles plays a crucial role in understanding spectral properties. 
In conventional fermionic systems in dimensions higher than one, it is known that excitations are well understood in terms of 
fermionic quasiparticles in the Fermi liquid theory, where single-particle excitations are created from a dense distribution of 
the quasiparticles in the region enclosed by the well-defined Fermi surface, and the excitations are gapless on the Fermi surface. 
On the other hand, in conventional bosonic systems, elementary excitations are generated from Bose-Einstein condensates or ordered ground states, 
where the gapless points are determined by the ordering momentum. 
\par
In one-dimensional (1D) chains, 
the statistics of particles may seem to be less important than that in higher dimensions. 
Indeed, 1D hard-core boson models (or equivalently $S$=1/2 spin chains) 
can be mapped onto 1D spinless fermion (SF) models through the Jordan-Wigner (JW) transformation~\cite{LSM}, 
and bulk quantities are exactly the same in these models. 
However, as will be shown in this Letter, the statistics causes substantial differences 
in spectral features, such as dispersion relations and line shapes of spectral functions, even for 1D chains in the entire energy range, 
which naturally connect to low-energy properties studied 
in detail using the field theory \cite{Giamarchi,Imambekov,Khodas,DMRG}. 
\par
In addition, systems with repulsive interactions exhibit interesting spectral features in the high-energy regime. 
For the spin-1/2 Heisenberg chain, it has been found that the high-energy continuum originating from 2-string solutions 
of the Bethe ansatz~\cite{Bethe} has a substantial spectral weight in a magnetic field~\cite{Kohno}. 
Since the Bethe equation of the SF model can be formally written in the same form as that of the XXZ model~\cite{LSM,Bethe}, 
we expect that the 2-string solutions will also contribute significantly to spectral weights in the high-energy regime for the SF model. 
Besides, numerical calculations using the time-dependent density-matrix renormalization group (tDMRG) method 
have indicated the continuum in the energy and momentum regions of the 2-string solutions for the SF model~\cite{DMRG}. 
In this Letter, we confirm the high-energy continuum with a considerable spectral weight in the SF model 
by directly using the 2-string solutions. 
\par
{\it Model }- 
In this Letter, we consider the 1D SF model with the nearest-neighbor interaction: 
\begin{equation}
\begin{split}
{\cal H}_{SF}=&\sum_{i=1}^{L}\left[t\left(c^{\dagger}_{i+1}c_i+c^{\dagger}_ic_{i+1}\right)+V\left(n_{i+1}-\frac{1}{2}\right)\left(n_{i}-\frac{1}{2}\right)\right]\\
&+\mu\sum_{i=1}^{L}\left(n_{i}-\frac{1}{2}\right),\nonumber
\end{split}
\end{equation}
where $c_i$ is the annihilation operator of a fermion at site $i$, and $n_i$ is the number operator defined as $n_i\equiv c^{\dagger}_{i}c_i$. 
Here, $t$, $V$, $\mu$, and $L$ represent the transfer integral, interaction strength, chemical potential, and 
number of sites, respectively. We impose the periodic [antiperiodic] 
boundary condition for chains with an odd [even] number of particles in the ground state. 
\par
The 1D SF model can be mapped onto the 1D spin-1/2 XXZ model under the periodic boundary condition 
\begin{equation}
\begin{split}
{\cal H}_{XXZ}=&\sum_{i=1}^{L}\left[\frac{J_{xy}}{2}\left(S^{+}_{i+1}S^{-}_i+S^{+}_iS^{-}_{i+1}\right)+J_zS^z_{i+1}S^z_{i}\right]\\
&-H\sum_{i=1}^{L}S^z_{i}\nonumber
\end{split}
\end{equation}
through the JW transformation~\cite{LSM} 
\begin{equation}
\begin{split}
&S_j^+=c_je^{-\i\pi\phi_j},\quad S_j^-=e^{\i\pi\phi_j}c^\dagger_j,\quad\mbox{and}\\
&S_j^z=\frac{1}{2}-n_j\quad\mbox{with}\quad\phi_j=\sum_{l=1}^{j-1}n_l.\nonumber
\end{split}
\end{equation}
In this transformation, the fermions in the SF model are mapped onto the down spins in the XXZ model with $J_{xy}=2t$, $J_z=V$, and $H=\mu$. 
Hereafter, the number of fermions in the SF model (or equivalently the number of down spins in the XXZ model) is denoted by $M$, 
and we assume that $L$ is even. 
In this Letter, we focus our attention on the case of $V/t=2$, which corresponds to the Heisenberg chain in a magnetic field ($J$=$J_{xy}$=$J_z$). 
\par
\begin{figure}[h]
\begin{center}
\includegraphics[width=5.98cm]{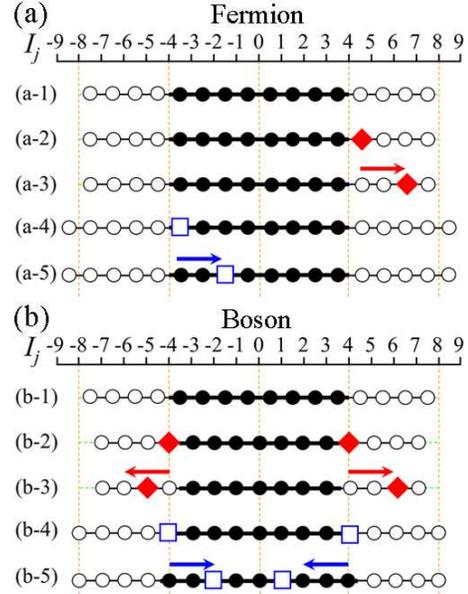}
\end{center}
\caption{(Color online) Typical distributions of Bethe quantum numbers $\{I_j\}$ in $L=24$ for (a) the spinless fermion model and (b) the Heisenberg model. 
Solid circles and red solid diamonds denote $I_j$ ($j=1\sim M$). Open circles and blue open squares denote unoccupied regions of $\{I_j\}$. 
Blue open squares and red solid diamonds behave as quasiparticles called psinons ($\psi$'s) and antipsinons ($\psi^*$'s), respectively \cite{Karbach_psinon}. 
The top rows show the distributions of $\{I_j\}$ in the ground states with $M$=8. 
The second rows are those of the lowest excited states with $M$=9 for $0\le k\le\pi$. 
The third rows are those of dynamically dominant excitations with $M$=9. 
The fourth rows are those of the lowest excited states with $M$=7 for $0\le k\le\pi$. 
The fifth rows are those of dynamically dominant excitations with $M$=7.}
\label{fig:Ij}
\end{figure}

{\it Method }- 
The eigenstates of the 1D SF model can be obtained using the Bethe ansatz~\cite{Bethe}. 
The wavefunctions for $V/t=2$ are expressed in terms of rapidities $\Lambda_j$ ($j=1\sim M$) that satisfy the following Bethe equation: 
\begin{equation}
L\arctan(\Lambda_j)=\pi I_j+\sum_{l=1}^M\arctan\left(\frac{\Lambda_j-\Lambda_l}{2}\right),
\label{eq:BE}
\end{equation}
where $I_j$ ($j=1\sim M$) are called Bethe quantum numbers. 
The difference in the statistics of particles is reflected in these quantum numbers. 
For the SF model, they are integers [half-odd integers] under the periodic [antiperiodic] boundary condition. 
On the other hand, for the Heisenberg model, they are integers [half-odd integers] under the periodic boundary condition in chains 
with an odd [even] number of down spins. 
Once a set of $\{I_j\}$ is given, a set of $\{\Lambda_j\}$ is obtained through eq.~(\ref{eq:BE}). 
Thus, eigenstates are characterized by distributions of $\{I_j\}$. Distributions of $\{I_j\}$ are somewhat analogous to 
momentum distributions of noninteracting fermions. Namely, $\{I_j\}$ in the ground state are consecutively distributed around zero 
like a Fermi sea, as shown in Figs. \ref{fig:Ij}(a-1) and \ref{fig:Ij}(b-1). Excitations from the ground state can be obtained by creating holes [particles] 
inside [outside] the consecutive distribution, as indicated by blue open squares [red solid diamonds] in Fig.~\ref{fig:Ij}. 
The hole and the particle are called psinon and antipsinon, 
and denoted by $\psi$ and $\psi^*$, respectively~\cite{Karbach_psinon}. 
\par
The momentum $K$ and the energy $E$ can be expressed in terms of $\{I_j\}$ or $\{\Lambda_j\}$ as 
\begin{equation}
K=M\pi-\frac{2\pi}{L}\sum_{j=1}^MI_j \quad(\mbox{mod}\quad2\pi), \\
\label{eq:k}
\end{equation}
\begin{equation}
E=-t\sum_{j=1}^M\frac{4}{\Lambda_j^2+1}+\frac{tL}{2}+\mu\left(M-\frac{L}{2}\right). 
\label{eq:E}
\end{equation}
The spectral functions $A^{\pm}(k,\omega)$ defined below~\cite{Akw} can also be expressed in terms of 
$\{\Lambda_j\}$~\cite{Motegi}. 
\begin{equation}
\begin{split}
&A^{+}(k,\omega)=\sum_{i}\left|\langle k,\epsilon_i|c^{\dagger}_{k}|\mbox{GS}\rangle\right|^2\delta(\omega-\epsilon_i),\\
&A^{-}(k,\omega)=\sum_{i}\left|\langle k,\epsilon_i|c_{-k}|\mbox{GS}\rangle\right|^2\delta(\omega+\epsilon_i),
\end{split}
\label{eq:Akw}
\end{equation}
where $|\mbox{GS}\rangle$ and $|k,\epsilon_i\rangle$ denote the ground state and the excited state 
with the excitation energy $\epsilon_i$ and momentum $k$ measured from those of the ground state. 
The spectral functions can be probed by angle-resolved photoemission experiments. 
The $A^{\pm}(k,\omega)$ correspond to dynamical structure factors $S^{\pm\mp}(k,\omega)$ 
for the XXZ model \cite{Kitanine,Kohno,Biegel,BiegelXXZ,Caux,Hagemans, Karbach_psinon}, 
where $c^{\dagger}_k$ and $c_{-k}$ in eq.~(\ref{eq:Akw}) are replaced by $S^-_k$ and $S^+_k$, respectively, 
and $S^{-+}(k,\omega)$ is reversed with respect to $\omega$=0 from that in the conventional definition with $\delta(\omega-\epsilon_i)$. 
Note that, even though wavefunctions have the same form, 
$|\langle k,\epsilon_i|c^{\dagger}_{k}|\mbox{GS}\rangle|^2\ne|\langle k,\epsilon_i|S^-_{k}|\mbox{GS}\rangle|^2$ because of the difference 
in the statistics~\cite{Motegi}. 
Meanwhile, 
$|\langle k,\epsilon_i|(n_{k}-1/2)|\mbox{GS}\rangle|^2=|\langle k,\epsilon_i|S^z_{k}|\mbox{GS}\rangle|^2$ \cite{Kitanine,Kohno,Biegel,BiegelXXZ,Caux,Hagemans, Karbach_psinon,Karbach_Szz,Sato}, because the number operator is bosonic. 
\par
We calculated $A^{\pm}(k,\omega)$ using the Bethe ansatz solutions 
with real $\Lambda_j$ ($j=1\sim M$) of up to 2$\psi$2$\psi^*$ excitations. 
For $A^+(k,\omega)$, we also considered solutions with a 2-string, i.e., a pair of complex rapidities having the same real part~\cite{Bethe,Takahashi}, 
with 1$\psi$1$\psi^*$ excitations for the real $\Lambda_j$ ($j=1\sim M-2$). 
We calculated $\mu$ in the thermodynamic limit~\cite{Griffiths} for $E$ in eq.~(\ref{eq:E}). 
\par
\begin{figure}[h]
\begin{center}
\includegraphics[width=8.7cm]{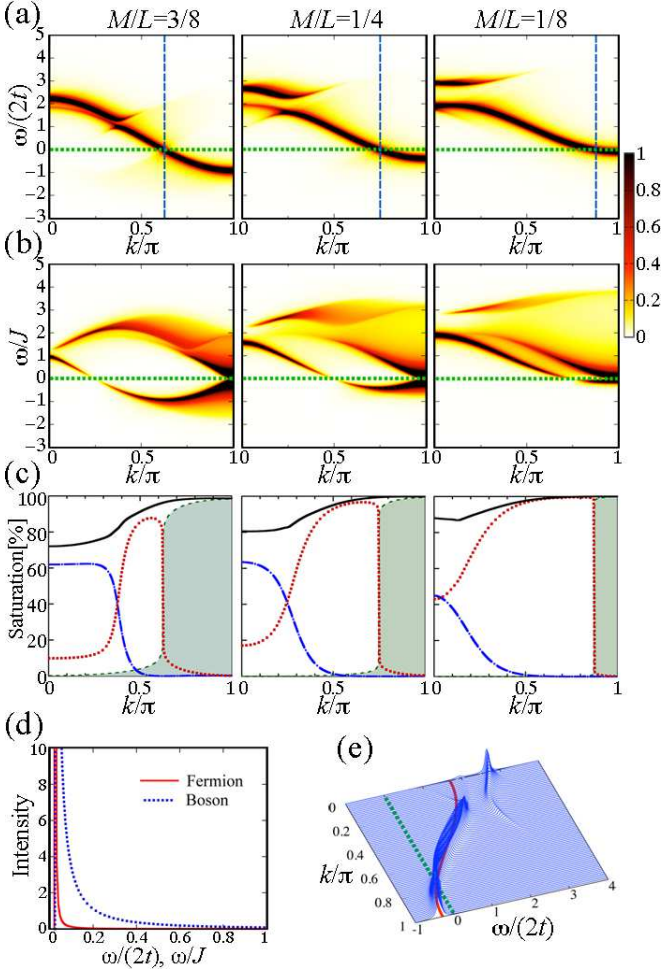}
\end{center}
\caption{(Color online) Spectral functions $A^{\pm}(k,\omega)$ of the spinless fermion (SF) model for $V/t$=2 and 
dynamical structure factors $S^{\pm\mp}(k,\omega)$ of the Heisenberg model ($J$=$J_{xy}$=$J_{z}$). 
Panels in (a), (b), and (c) show results in $L$=320 for $M/L$=3/8, 1/4, and 1/8 (from the left), respectively. 
(a) $A^{\pm}(k,\omega)$ of the SF model. 
(b) $S^{\pm\mp}(k,\omega)$ of the Heisenberg model~\cite{Kohno}. 
Here, $S^{-+}(k,\omega)$ are reversed with respect to $\omega$=0. 
The horizontal dotted lines in (a) and (b) denote the zero energy, and the vertical lines in (a) denote $k=\pi-k_F$. 
The data in (a) and (b) are broadened in a Lorentzian form with full width at half maximum 0.08$J$. 
(c) Contributions to the sum rule for the SF model. 
Green hatched regions denote the contributions from the real $\{\Lambda_j\}$ solutions of up to 2$\psi$2$\psi^*$ excitations for $A^-(k,\omega)$. 
Red dotted lines indicate those for  $A^+(k,\omega)$. 
Blue dashed-dotted lines show contributions of 2-string solutions with 1$\psi$1$\psi^*$ excitations for $A^+(k,\omega)$. 
Black solid lines indicate the sum of such contributions. (d) Line shapes of $A^{+}(k,\omega)$ and  $S^{+-}(k,\omega)$ near the gapless points for $M/L$=3/8. 
The red solid line denotes the line shape of 2$\psi^*$ excitations in the SF model at $k\simeq\pi-k_F=5\pi/8$, 
and the blue dotted line denotes that in the Heisenberg model at $k\simeq\pi$ in $L$=2240. 
(e) $A^{\pm}(k,\omega)$ of the SF model for $M/L$=1/4 in $L$=320. The green dotted line denotes the zero energy, 
and the red solid line shows the dispersion relation of the noninteracting system ($V$=0).}
\label{fig:Akw}
\end{figure}
{\it Comparisons between fermionic and bosonic systems }-
The results for the spectral functions $A^{\pm}(k,\omega)$ in the SF model 
and those for dynamical structure factors $S^{\pm\mp}(k,\omega)$ in the Heisenberg model~\cite{Kohno} are shown in Figs. \ref{fig:Akw}(a) and \ref{fig:Akw}(b), respectively. 
One of the differences is the position of the gapless points. 
In the SF model, the gapless points are located at $k=\pi\pm k_F$ with the Fermi momentum $k_F=\pi M/L$ as denoted by the vertical dashed lines in Fig.~\ref{fig:Akw}(a), 
whereas the gapless point at $k=\pi$ does not shift as a function of $M$ or the magnetization ($S^z=L/2-M$) in the Heisenberg model [Fig.~\ref{fig:Akw}(b)]. 
Another difference appears in the line shapes of $A^{\pm}(k,\omega)$ and $S^{\pm\mp}(k,\omega)$. 
Typical behaviors near the gapless points are shown in Fig.~\ref{fig:Akw}(d). 
In the SF model, the peak is very sharp and looks almost $\delta$-functional, as shown by the red solid line in Fig.~\ref{fig:Akw}(d), 
while that in the Heisenberg model has a larger tail in a wide range of energies, as shown by the blue dotted line in Fig.~\ref{fig:Akw}(d). 
These differences originate from distributions of $\{I_j\}$ that reflect the statistics of particles, as shown below. 
\par
{\it Gapless points }-
In the SF model, when $\{I_j\}$ in the ground state are half-odd integers [integers], 
those in excited states with one fermion added or removed are also half-odd integers [integers], as shown in Fig. \ref{fig:Ij}(a). 
Hence, the lowest excited state contributing to $A^+(k,\omega)$ has the distribution of $\{I_j\}$ with one $\psi^*$ attached to the right [left] edge of 
the consecutive distribution in the ground state as in Fig. \ref{fig:Ij}(a-2), which has the momentum $k$=$\pi$$-$$k_F$ [$\pi$+$k_F$] measured from that of the ground state (eq.~(\ref{eq:k})). 
Similarly, the lowest excited state ($\epsilon_i$$\rightarrow$+0) in $A^-(k,\omega)$ has the distribution of $\{I_j\}$ with one $\psi$ created 
at the left [right] edge of the consecutive distribution in the ground state as in Fig. \ref{fig:Ij}(a-4), which has $k$=$\pi$$-$$k_F$ [$\pi$+$k_F$]. 
Thus, the gapless points are located at $k$=$\pi$$\pm$$k_F$ for the SF model as in the noninteracting case ($V=0$). 
\par
On the other hand, in the bosonic model, $\{I_j\}$ in excited states with one down spin added or removed are 
integers [half-odd integers], when those in the ground state are half-odd integers [integers] as in Fig. \ref{fig:Ij}(b). 
Hence, the lowest excited state ($\epsilon_i$$\rightarrow$+0) in $S^{+-}(k,\omega)$ and that in $S^{-+}(k,\omega)$ have 
$\{I_j\}$ symmetrically distributed around zero as in Figs. \ref{fig:Ij}(b-2) and \ref{fig:Ij}(b-4). 
As a result, they have the momentum $k$=$\pi$ measured from that of the ground state (eq. (\ref{eq:k})) regardless of the magnetization value. 
\par
Thus, the gapless points differ, depending on the statistics of particles. 
The similarity between excitation spectra in the fermionic system (Fig.~\ref{fig:Akw}(a)) and those in the bosonic system (Fig.~\ref{fig:Akw}(b)) may be more easily seen 
by reversing $A^-(k,\omega)$ ($\omega<0$) in Fig.~\ref{fig:Akw}(a) with respect to $k=\pi-k_F$ (the vertical dashed lines in Fig.~\ref{fig:Akw}(a)) 
and shifting momenta in Fig.~\ref{fig:Akw}(a) by $k_F$ to the right. 
\par
{\it Dynamically dominant excitations }-
More interestingly, the above difference in $\{I_j\}$ distribution between the fermionic and bosonic systems also causes the difference 
in line shape between $A^{\pm}(k,\omega)$ and $S^{\pm\mp}(k,\omega)$. 
In general, excited states with distributions of $\{I_j\}$ close to that of the ground state carry large spectral weights. 
Thus, dynamically dominant excitations for $A^+(k,\omega)$ in the SF model are obtained by shifting $\psi^*$ 
in the lowest excited state (the red solid diamond in Fig.~\ref{fig:Ij}(a-2)) away from the center, as indicated by the red arrow in Fig.~\ref{fig:Ij}(a-3). 
These excitations appear in the momentum region $|k|\lesssim\pi-k_F$, 
as shown in Fig.~\ref{fig:Akw}(a) for $\omega\gtrsim0$. 
Similarly, dynamically dominant excitations for $A^-(k,\omega)$ are obtained by shifting $\psi$ 
in the lowest excited state (the blue open square in Fig.~\ref{fig:Ij}(a-4)) toward the center, 
as indicated by the blue arrow in Fig.~\ref{fig:Ij}(a-5). These excitations appear in the momentum region $|k\pm\pi|<k_F$, as shown in Fig.~\ref{fig:Akw}(a) for $\omega<0$. 
\par
These behaviors are in contrast to those in the bosonic model. 
Because of the symmetric distribution of $\{I_j\}$ in the lowest excited state for $S^{+-}(k,\omega)$ (Fig.~\ref{fig:Ij}(b-2)), 
dynamically dominant excitations are obtained by shifting 2 $\psi^*$'s (the red solid diamonds in Fig.~\ref{fig:Ij}(b-2)) 
away from the center as in Fig.~\ref{fig:Ij}(b-3). 
These excitations form the low-energy continuum near $k=\pi$ in $S^{+-}(k,\omega)$, as shown in Fig.~\ref{fig:Akw}(b) for $\omega>0$~\cite{Kohno}. 
Similarly, because the lowest excited state ($\epsilon_i\rightarrow+0$) in $S^{-+}(k,\omega)$ has 
the symmetric distribution of $\{I_j\}$ around zero (Fig.~\ref{fig:Ij}(b-4)), 
dynamically dominant excitations are obtained by shifting 2 $\psi$'s (the blue open squares in Fig.~\ref{fig:Ij}(b-4)) 
toward the center as in Fig.~\ref{fig:Ij}(b-5)~\cite{Karbach_psinon}. 
These excitations form the continuum near $k=\pi$ for $\omega<0$, as shown in Fig.~\ref{fig:Akw}(b)~\cite{Kohno,Karbach_psinon}. 
\par
As seen above, dynamically dominant excitations are different between the fermionic and bosonic systems 
owing to the difference in $\{I_j\}$ distribution. 
In the SF model, large spectral weights are carried by the 1$\psi^*$ [1$\psi$] mode with a small tail~\cite{Sakai} 
originating from less dominant multi-$\psi\psi^*$ excitations, as shown in Figs. \ref{fig:Akw}(a) and \ref{fig:Akw}(d). 
In contrast, in the bosonic model, spectral weights spread in the 2$\psi^*$ [2$\psi$] continuum, which shows a peak with a larger tail, 
as shown in Figs. \ref{fig:Akw}(b) and \ref{fig:Akw}(d). 
In this sense, $\psi^*$'s [$\psi$'s] may appear more evenly fractionalized in bosonic systems than in fermionic systems. 
\par
{\it High-energy continuum }-
Next, we discuss the high-energy continuum in Fig. \ref{fig:Akw}(a). 
For the Heisenberg model, it has been found that the continuum of the 2-string solutions has a large spectral weight in $S^{+-}(k,\omega)$~\cite{Kohno}, 
which has resolved a long-standing puzzle on the high-energy properties of quasi-1D antiferromagnets in a magnetic field~\cite{Kohno,Kohno_q1DH}. 
Because the Bethe equation of the SF model can be formally written in the same form as that of the Heisenberg model (eq.~(\ref{eq:BE})), 
it is naturally expected that the 2-string solutions in the SF model will also have considerable spectral weights in the high-energy regime. 
Besides, numerical calculations using the tDMRG method have indicated the continuum 
in the energy and momentum regions of the 2-string solutions for the SF model~\cite{DMRG}. 
We directly confirmed it by calculating $A^{\pm}(k,\omega)$ using the $\{\Lambda_j\}$ of the 2-string solutions for the SF model. 
Although the low-energy continuum from real $\{\Lambda_j\}$ solutions remains almost on the cosine dispersion relation 
of the noninteracting system (the red solid line in Fig.~\ref{fig:Akw}(e)), the high-energy continuum of the 2-string solutions is separated from it, 
as shown in Figs. \ref{fig:Akw}(a) and \ref{fig:Akw}(e). 
\par
We also investigated their contribution to the sum rule, 
\begin{equation}
\begin{split}
1&=\int_{-\infty}^{\infty}d\omega\left[A^+(k,\omega)+A^-(k,\omega)\right]\\
&=\sum_i\left[|\langle k,\epsilon_i|c^{\dagger}_k|\mbox{GS}\rangle|^2+|\langle k,\epsilon_i|c_{-k}|\mbox{GS}\rangle|^2\right],\nonumber
\end{split}
\end{equation}
as shown in Fig.~\ref{fig:Akw}(c). The contribution of the 2-string solutions is large near $k=0$. 
Indeed, they occupy more than 60\% of the total spectral weight per momentum near $k=0$ for $M/L$=3/8 and 1/4. 
The missing weight would be mainly accounted for by 3-string solutions as in the Heisenberg model~\cite{Kohno}. 
It should be noted that, in the SF model, the momenta of the 2-string solutions are shifted by $k_F(=\pi M/L)$ from those of the Heisenberg model, 
because $\{I_j\}$ for real $\Lambda_j$ ($j=1\sim M-2$) in the 2-string solutions are integers [half-odd integers] 
when those in the ground state are integers [half-odd integers] as in the case of real $\{\Lambda_j\}$ solutions discussed above. 
\par
We expect that such a high-energy continuum with a substantial spectral weight will appear in more general 1D systems with repulsive interactions. 
In preliminary calculations on the SF model with $V>0$ and the XXZ model with $J_z>0$, 
we have confirmed that the high-energy continuum of the 2-string solutions has a considerable spectral weight 
in both weak-coupling (XY-like) and strong-coupling (Ising-like) regimes. 
The results will be presented elsewhere. 
Also, it should be noted that solutions with a pair of fermions in the 1D Hubbard model have energies of $O(U)$ in the large-$U$ regime, 
corresponding to the upper Hubbard band~\cite{HubbardBand}. 
\par
{\it Summary }-
We have investigated the one-particle spectral functions of the 1D spinless fermion model with the nearest-neighbor repulsion using Bethe ansatz solutions. 
The main message is that the statistics of particles substantially affects dynamical features, such as the gapless points 
and line shapes of spectral functions, even in 1D chains. 
In the SF model, the distributions of Bethe quantum numbers $\{I_j\}$ in dynamically dominant excitations for $A^+(k,\omega)$ [$A^-(k,\omega)$] 
have one extra $\psi^*$ [$\psi$], which results in the gapless points at $k=\pi\pm k_F$ with $k_F=\pi M/L$ and sharp peaks in $A^{\pm}(k,\omega)$. 
In contrast, in the bosonic model, 
because of the symmetric distribution of $\{I_j\}$ in the lowest excited states, 2$\psi^*$ [2$\psi$] excitations are dynamically dominant. 
This results in the gapless point at $k=\pi$ regardless of the magnetization value and gives rise to a larger tail in $S^{\pm\mp}(k,\omega)$ 
than in $A^{\pm}(k,\omega)$. 
\par
We also confirmed the high-energy continuum with a considerable spectral weight in the SF model by directly using the 2-string solutions. 
We expect that such a high-energy continuum will be a general feature in 1D systems with repulsive interactions. 
The arguments in this Letter will be applicable to more general 1D fermionic and bosonic systems with repulsive interactions. 
\section*{Acknowledgement} 
We are grateful to K. Motegi, M. Shiroishi, and M. Takahashi for discussions and helpful comments. This work was supported by 
World Premier International Research Center Initiative on Materials Nanoarchitectonics, 
the University of Tsukuba Research Initiative, and KAKENHI 18340112, 20029004, 20046002, 20046015, 20654034, 20740206, 21740281, and 21740285.


\begin{thebibliography}{9}
\bibitem{LSM} E. Lieb, T. Schultz, and D. Mattis: Ann. Phys. (N.Y.) {\bf 16} (1961) 407, and references therein.
\bibitem{Giamarchi} T. Giamarchi: {\it Quantum Physics in One Dimension}, 
Vol. 121 of {\it International Series of Monographs on Physics} (Oxford Science Publications, Clarendon Press, Oxford, 2003).
\bibitem{Imambekov} A. Imambekov, and L. I. Glazman: Science {\bf 323} (2009) 228.
\bibitem{Khodas} M. Khodas, M. Pustilnik, A. Kamenev, and L. I. Glazman: Phys. Rev. B {\bf 76} (2007) 155402.
\bibitem{DMRG} R. G. Pereira, S. R. White, and I. Affleck: Phys. Rev. B {\bf 79} (2009) 165113.
\bibitem{Bethe} H. Bethe: Z. Phys. {\bf 71} (1931) 205.
\bibitem{Kohno} M. Kohno: Phys. Rev. Lett. {\bf 102} (2009) 037203.
\bibitem{Karbach_psinon} M. Karbach, D. Biegel, and G. M\"uller: Phys. Rev. B {\bf 66} (2002) 054405.
\bibitem{Akw} The $A^-(k,\omega)$ defined here is typically denoted as $A^-(-k,\omega)$. 
In the models with inversion symmetry, $A^{\pm}(k,\omega)$=$A^{\pm}(-k,\omega)$. 
\bibitem{Motegi} K. Motegi, and K. Sakai: Nucl. Phys. B {\bf 793} (2008) 451.
\bibitem{Kitanine} N. Kitanine, J. M. Maillet, and V. Terras: Nucl. Phys. B {\bf 554} (1999) 647.
\bibitem{Biegel} D. Biegel, M. Karbach, and G. M\"uller: Europhys. Lett. {\bf 59} (2002) 882.
\bibitem{BiegelXXZ} D. Biegel, M. Karbach, and G. M\"uller: J. Phys. A: Math. Gen. {\bf 36} (2003) 5361.
\bibitem{Caux} J. -S. Caux, R. Hagemans, and J. M. Maillet: J. Stat. Mech. (2005) P09003.
\bibitem{Hagemans} R. Hagemans, J. -S. Caux, and J. M. Maillet: AIP Conf. Proc. {\bf 846} (2006) 245. 
\bibitem{Karbach_Szz} M. Karbach, and G. M\"uller: Phys. Rev. B {\bf 62} (2000) 14871.
\bibitem{Sato} J. Sato, M. Shiroishi, and M. Takahashi: J. Phys. Soc. Jpn.  {\bf 73} (2004) 3008. 
\bibitem{Takahashi} M. Takahashi: Prog. Theor. Phys. {\bf 46} (1971) 401.
\bibitem{Griffiths} R. B. Griffiths: Phys. Rev. {\bf 133} (1964) A768.
\bibitem{Sakai} K. Sakai, M. Shiroishi, J. Suzuki, and Y. Umeno: Phys. Rev. B {\bf 60} (1999) 5186.
\bibitem{Kohno_q1DH} M. Kohno: Phys. Rev. Lett. {\bf 103} (2009) 197203.
\bibitem{HubbardBand} M. Takahashi: Prog. Theor. Phys. {\bf 47} (1972) 69.
\end{thebibliography}
\end{document}